\documentclass[floatfix,twocolumn,aps,prd,showpacs,amsmath,amssymb]{revtex4}
\usepackage{bm}
\usepackage{epsfig}
\usepackage{psfrag}
\begin{document}


\title{Is the equivalence for the response of static scalar sources
       in the Schwarzschild and Rindler spacetimes valid only in 
       four dimensions?}

\author{Lu\'\i s C.\ B.\ Crispino}
\email{crispino@ufpa.br}
\affiliation{Departamento de F\'\i sica, Universidade Federal do Par\'a,
Campus Universit\'ario do Guam\'a, 66075-900, Bel\'em, Par\'a, Brazil}
\author{Atsushi Higuchi}
\email{ah28@york.ac.uk}
\affiliation{Department of Mathematics, University of York, Heslington, 
York YO10 5DD, United Kingdom}
\author{George E.\ A.\ Matsas}
\email{matsas@ift.unesp.br}
\affiliation{Instituto de F\'\i sica Te\'orica, Universidade Estadual Paulista, 
Rua Pamplona 145, 01405-900, S\~ao Paulo, SP, Brazil}
\date{\today}

\begin{abstract}
It was shown recently that in four dimensions 
scalar sources with fixed proper acceleration
minimally coupled to a massless Klein-Gordon field lead to 
the same responses when they are
(i) uniformly accelerated in Minkowski spacetime (in the inertial vacuum) and
(ii) static in the Schwarzschild spacetime (in the Unruh vacuum).
Here we show that this equivalence is broken if the spacetime dimension
is more than four.
\end{abstract}

\maketitle

 
Let us consider a pointlike scalar source with fixed proper acceleration, 
$a_0 = {\rm const}$,  minimally coupled to a massless Klein-Gordon 
field $\Phi$ through a small coupling constant $q$. 
It was shown recently that the source's response $R_{\rm S} (r_0,M)$ 
to the Hawking radiation (associated with the Unruh vacuum) obtained 
when it lies at rest with (Schwarzschild) radial coordinate 
$r_0 = {\rm const} > 2M$, outside a chargeless static black hole with mass
$M$, is exactly the same as the response $R_{\rm M} (a_0)$ of the source 
when it is uniformly accelerated (with the same proper acceleration as 
before) in the inertial vacuum of the Minkowski 
spacetime, or, equivalently, when it is static in the Fulling-Davies-Unruh 
thermal bath of the Rindler wedge~\cite{HMS2}.

The fact that this result is surprising can be seen as follows. First
let us recall that in Schwarzschild spacetime we can express the
source's radial coordinate $r_0$ in terms of its proper acceleration
$a_0$ and the black hole mass $M$: $r_0 = r_0 (a_0,M)$. Thus, it
would be natural to expect that the response would depend on $M$
as well as on $a_0$, {\em i.e.}, $R_{\rm S}= R_{\rm S} (a_0,M)$, 
rather than
\begin{equation}
R_{\rm S} = R_{\rm S} (a_0) = R_{\rm M} (a_0) = {q^2 a_0}/{(4 \pi^2)}\, . 
\label{equivalence}
\end{equation}
We note that structureless static scalar sources can only 
interact with zero-energy field modes. Such modes probe the global 
geometry of spacetime and are accordingly quite different in 
Schwarzschild and Rindler spacetimes. 
Indeed the equivalence~(\ref{equivalence}) is not valid, for instance, 
if one replaces the Unruh vacuum by the Hartle-Hawking one, 
in which case the source's response is~\cite{HMS2}
$$
R'_S (a_0,M) =  
             {q^2 a_0}/{(4 \pi^2)} + {q^2}/{(16 \pi^2 r_0^2 a_0)} \;, 
$$ 
nor when the massless Klein-Gordon field is replaced by 
electromagnetic~\cite{CHM1} or massive Klein-Gordon~\cite{CCSM} one. 
Moreover, the equivalence was shown to be broken
also when the background 
spacetime is endowed with a cosmological constant~\cite{CCSM2} or when 
the black hole is given some electric charge~\cite{CM}. 

It is hitherto unclear whether or not the equivalence found in 
Ref.~\cite{HMS2} hides something deeper behind it. Even in the less 
interesting case where the 
equivalence turns out to be a ``coincidence'', it will still 
be interesting to determine 
whether or not this is precisely restricted to the number of 
(macroscopic) dimensions of physical spacetimes, as we will do 
in this paper.
Here we adopt natural units ($c=G=\hbar=k_B=1$) 
and spacetime signature $(+-\dots-)$.

The line element of a Schwarzschild spacetime with $N \equiv p+2$ 
dimensions ($p \geq 2$) is~\cite{MP}
$$
ds^2 = f(r) dt^2 - f(r)^{-1} dr^2 - r^2 ds^2_p \; ,
$$
where
$
f(r) = 1- (r_H/r)^{p-1} 
$ with
$r_H = (2M)^{1/(p-1)}$ being the radius of the event horizon 
and $ds^2_p$ being the line element of a unit $p$-sphere $S^p$. 
This is assumed to be covered with angular coordinates 
$\{\theta_1, \dots, \theta_p \}$ and to be endowed with the 
standard metric  $\tilde \eta_{ij}$ (and inverse metric 
$\tilde \eta^{ij}$) with signature $(+\dots+)$. Here 
$i= 1, \dots, p$ and $j= 1, \dots, p$ are associated 
with the angular coordinates on $S^p$. 

The Klein-Gordon equation $\Box \Phi = 0$ can be written in 
this background as
\begin{equation}
{f}^{-1} \partial_t^2 \Phi - r^{-p} \partial_r 
\left[ f r^p \partial_r \Phi \right] - 
r^{-2} {\widetilde \nabla}^2 \Phi =0  \, ,
\label{KGequation}
\end{equation}
where 
$
{\widetilde \nabla}^2 \equiv 
\tilde \eta^{ij} {\widetilde \nabla}_i {\widetilde \nabla}_j$
is the Laplacian and ${\widetilde \nabla}_i $ is the 
associated covariant derivative on $S^p$.
We look for positive frequency modes in the form 
\begin{equation}
u^{(n)}_{\omega l m} (t,r,\theta_i) = 
      \psi^{(n)}_{\omega l}(r) Y_{lm}(\theta_i) e^{-i\omega t} 
\label{modes}
\end{equation}
associated with the timelike Killing field 
$ \xi = \partial / \partial t$, where
$\omega \geq 0$, 
$n= \rightarrow$ and $\leftarrow$ label purely ingoing modes from 
the past white hole horizon ${\cal H}^-$ and from the 
past null infinity ${\cal J}^-$, respectively, and 
$l=0,1,2,\dots$, and $m$ denotes a set 
$\{ m_1, \dots, m_{p-1} \}$ of $p-1$ integers satisfying 
$l\geq m_{p-1}\geq \dots \geq m_2 \geq | m_1 |$. 
The $u^{(n)}_{\omega l m}(x^\mu)$ modes are assumed to be orthonormalized 
with respect to the Klein-Gordon inner product~\cite{BD}:
\begin{eqnarray}
\label{inner}
     i&\int_{\Sigma_t} & d\Sigma^{p+1} \; n^\mu 
     \left( 
     {u^{(n)}_{\omega l m}}^* \nabla_\mu u^{(n')}_{\omega' l' m'} - 
     u^{(n')}_{\omega' l' m'} \nabla_\mu {u^{(n)}_{\omega l m}}^*  
     \right) \nonumber \\ 
     &=& 
     \delta_{n n'} \delta_{l l'} 
     \delta_{m m'} \delta(\omega - \omega') \;,
     \\
     i&\int_{\Sigma_t}& d\Sigma^{p+1} \;  n^\mu 
     \left( 
     u^{(n)}_{\omega l m} \nabla_\mu u^{(n')}_{\omega' l' m'} 
     - u^{(n')}_{\omega' l' m'} \nabla_\mu u^{(n)}_{\omega l m}  
     \right)\nonumber \\  
     &=&
     0 \;,\nonumber
\end{eqnarray}
where $n^\mu$ is the future-directed unit vector normal to 
the Cauchy surface $\Sigma_t$, {\it e.g.}, $t={\rm const}$. 
We note that modes $n= \rightarrow$ and $\leftarrow$ 
are orthogonal to each other. This fact can easily 
be seen by choosing 
$\Sigma_t ={\cal H}^- \cup {\cal J}^- $ in Eq.~(\ref{inner}) 
and recalling that $\psi_{\omega l}^{(\rightarrow)}(x)$ 
and $\psi_{\omega l}^{(\leftarrow)}(x) $ 
vanish on $ {\cal J}^-$ and ${\cal H}^-$, respectively.

The modes $u^{(n)}_{\omega l m}$ and their respective 
complex conjugates form a complete orthonormal basis in the space 
of solutions of Eq.~(\ref{KGequation}). As a result, we can expand 
the field  operator as 
\begin{equation}
 \label{expansion}
     \hat{\Phi}(x^\mu) = 
        \sum_{n=\leftarrow}^{\rightarrow} 
              \sum_{l=0}^{\infty} \sum_m
        \int_0^{\infty} d\omega 
        \left[ 
        u^{(n)}_{\omega l m} \hat{a}^{(n)}_{\omega l m} + {\rm H.c.} 
        \right] \;,
  \end{equation}
where $\hat{a}^{(n)}_{\omega l m} $ and $\hat{a}^{(n) \dagger }_{\omega l m} $ 
are annihilation and creation operators, respectively, and satisfy the usual 
commutation relations
$
[ \hat{a}^{(n)}_{\omega l m}, \hat{a}^{(n') \dagger }_{\omega' l' m'} ] = 
    \delta_{n n'} \delta_{l l'}
    \delta_{m m'}\delta(\omega - \omega')
$.   
The Boulware vacuum $|0\rangle$ is defined by
$\hat{a}^{(n)}_{\omega lm}|0\rangle = 0$ for all $n, \omega, l$ and
$m$. This is the state of ``no particles'' as defined by the static
observers following integral curves of the vector field 
$\xi = \partial/\partial t$. 

Next, by substituting
Eq.~(\ref{modes}) in the Klein-Gordon equation and using
$\widetilde \nabla^2 Y_{lm} = -l(l+p-1) Y_{lm}$ (for  
spherical harmonics on $p$-spheres see, {\it e.g.}, 
Ref.~\cite{CM2}), we obtain 
\begin{equation}
\frac{f}{r^p} \frac{d}{dr} 
\left( r^p f \frac{d\psi^{(n)}_{\omega l}}{dr} \right)
+ \left[ \omega^2 -l(l+p-1)\frac{f}{r^2} \right] 
\psi^{(n)}_{\omega l} = 0 \, .
\label{Rwdif0}
\end{equation}
Now we define 
$\varphi^{(n)}_{\omega l}(r) \equiv r^{p/2} \psi^{(n)}_{\omega l}(r)$
and 
$d/dx \equiv f(r) \, d/dr$ to cast Eq.~(\ref{Rwdif0}) in the form
\begin{equation}
[ {d^2}/{dx^2} + \omega^2 - V_{\rm eff}(x) ] 
\varphi^{(n)}_{\omega l}(r) = 0 \, ,
\label{Eqinx}
\end{equation}
where the scattering potential is
$$
V_{\rm eff}[x(r)] = f\left[ \frac{pf'}{2r} 
+ \frac{p}{2} \left( \frac{p}{2} -1 \right) \frac{f}{r^2}
+ \frac{l(l+p-1)}{r^2}
\right]
$$
with
$f' \equiv df/dr$ 
and
\begin{equation}
x(r) = \left\{
\begin{array}{ll}
r + r_H \ln (r/r_H -1 ) 
\; & {\rm for} \; p = 2  
\\
r \; F \left[ \frac{1}{1-p},1;\frac{2-p}{1-p}; 
                  \left( \frac{r_H}{r} \right)^{p-1} \right]
\; & {\rm for} \; p\geq 3
\end{array}
\right. \, .
\label{x(r)}
\end{equation}

Close  ($x<0, |x| \gg r_H $) to and far away ($x \gg r_H $) 
from the horizon, we have $V_{\rm eff} [x(r)] \approx 0$, and we write 
\begin{equation}
     \varphi_{\omega l}^{(\rightarrow)} \! \approx \!
        \left\{ 
          \begin{array}{lc}
         A^{(\rightarrow)}_{\omega l} \left( e^{i\omega x} + 
         {\cal R}_{\omega l}^{(\rightarrow)} 
         e^{-i\omega x} \right) & (x < 0\;, |x| \gg r_H)  \\
         A^{(\rightarrow)}_{\omega l}   
          {\cal T}_{\omega l}^{(\rightarrow)} e^{i\omega x} 
          & (x \gg r_H) 
          \end{array} 
        \right.   
	\label{51a}
\end{equation}
and
\begin{equation}
     \varphi_{\omega l}^{(\leftarrow)} \! \approx \!
        \left\{ 
          \begin{array}{lc}
            A^{(\leftarrow)}_{\omega l} 
            {\cal T}_{\omega l}^{(\leftarrow)} e^{-i\omega x} 
            & (x< 0, |x| \gg r_H) \\
            A^{(\leftarrow)}_{\omega l} 
            ( e^{-i\omega x} + {\cal R}_{\omega l}^{(\leftarrow)}e^{i\omega x}) 
            & (x \gg r_H).
          \end{array} \right. 
	\label{51b}
\end{equation}
Here 
$
| {\cal R}_{\omega l}^{(n)} |^2 
$  
and
$
| {\cal T}_{\omega l}^{(n)} |^2 
$
are the reflection and transmission coefficients, respectively, 
satisfying the usual probability conservation equation 
$
| {\cal R}_{\omega l}^{(n)} |^2 
+ 
| {\cal T}_{\omega l}^{(n)} |^2 
= 1 
$. 
The normalization constants $A^{(n)} _{\omega l}$ can be obtained 
from the 
Klein-Gordon inner product~(\ref{inner}), which implies 
$$
\int^{+\infty}_{-\infty} dx
\varphi^{(n) *}_{\omega l}(r) \varphi^{(n')}_{\omega' l}(r) =
\delta_{n n'}(\omega +\omega')^{-1} \delta (\omega - \omega') \;.
$$
In order to transform the integral into a surface term
(see~\cite{HMS2} for more details in four dimensions), we use
Eq.~(\ref{Eqinx}) in addition to
$
| {\cal R}_{\omega l}^{(n)} |^2 
+ 
| {\cal T}_{\omega l}^{(n)} |^2 
= 1 \; 
$,
which leads (up to an arbitrary phase) to 
$A^{(n)}_{ \omega l} = 1/(2\sqrt{\pi \omega})$.

Let us now describe our pointlike  scalar source lying 
at $(r_0, {\theta_{i}}_0)$ by
\begin{equation}
     j(x^\mu) = ({q}/{\sqrt{|h|\;}}) 
                \delta(r-r_0) 
                \delta^p(\theta_i - {\theta_{i}}_0 ) \;,
     \label{j(x)}
\end{equation}
where we recall that $q$ is a small constant and 
$h= {\rm det} (g_{ij}) $ 
is the determinant of the spatial 
metric on $\Sigma_t$. Note that 
$
     \int_{\Sigma_t} d{\Sigma}^{p+1} \; j(x^\mu) = q 
$
wherever the source lies. The  absolute value
of the source's four-acceleration  
$a_0= | u^{\mu} \nabla_{\mu} u^{\nu} |$ 
is 
\begin{equation}
a_0 = \frac{(p-1) r_H^{p-1}}{2 r_0^p \sqrt{1-(r_H/r_0)^{p-1}}} \; ,
\label{4acceleration}
\end{equation}
where we have used
$
u^{\mu} = f^{-1/2}(r_0)\delta^{\mu}_t \, .
$

Now, let us couple our scalar source $j(x^\mu)$ to the 
Klein-Gordon field $\hat \Phi (x^\mu) $ as described by the 
interaction action
\begin{equation}
     \hat S_I = 
       \int dx^{p+2} \sqrt{|g|}\; j(x^\mu )\; \hat \Phi (x^\mu ) \;.
  \label{SI}
\end{equation}

The total source response, {\it i.e.}, total particle emission and
absorption probabilities per proper time of the source,
is given in a thermal bath by
\begin{equation}
     R_S \equiv  
       \sum_{n=\leftarrow}^{\rightarrow} 
       \sum_{l=0}^{\infty} 
       \sum_{m} 
       \int_0^{\infty} d\omega  R^{(n)}_{\omega l m}  \;  ,
   \label{totalresponsegeneral}
\end{equation}
where 
\begin{equation}
     R^{(n)}_{\omega lm} \equiv 
     \frac{1}{\tau}
       \left\{  
     |{{\cal A}^{(n){\rm em}}_{\omega lm}}  |^2 
      [1+ n^{(n)}(\omega)] 
 +  
     |{{\cal A}^{(n){\rm abs}}_{\omega lm}} |^2 
      n^{(n)}(\omega) 
     \right\}
   \label{responsegeneral}
\end{equation}
and $\tau = 2\pi \sqrt{f(r_0)\,} \delta(0)$ is the  source's total 
proper time~\cite{HMS2}. Here
$
{{\cal A}^{(n){\rm em}}_{\omega lm}} 
\equiv 
\langle n \omega l m | \hat S_I | 0 \rangle
$
and
$
{{\cal A}^{(n){\rm abs}}_{\omega lm}}
\equiv 
\langle 0 | \hat S_I  | n \omega l m \rangle
$
are the emission and absorption amplitudes,   respectively,
of Boulware states  $|n \omega lm\rangle$, at the tree level,
and 
\begin{equation}
n^{(n)}(\omega)
\equiv 
\left\{ 
 \begin{array}{cl}
 ( e^{\omega\beta } - 1 )^{-1} & \;\;\; {\rm for} \;\;\; n = \rightarrow \\
 0                             & \;\;\; {\rm for} \;\;\; n = \leftarrow 
 \end{array} 
\right.
\label{n},
\end{equation}
for the Unruh vacuum. We recall that the Unruh vacuum is 
characterized by a thermal flux 
leaving  ${\cal H}^-$ with Hawking temperature $\beta^{-1}$ at 
infinity and no thermal flux coming from  ${\cal J}^-$.
Here $\beta^{-1} = {\cal K}/(2\pi)$ as is well known~\cite{W} with the 
surface gravity 
$
{\cal K} = (p-1)/(2r_H)
$.
Since structureless static sources~(\ref{j(x)}) 
can only interact with {\em zero-energy} modes,  the total response 
of this  source  in the Boulware vacuum  vanishes (for a more 
comprehensive discussion on zero-energy modes, see Ref.~\cite{HMS}). 
This is not so, however, in the presence of a background thermal bath
since the absorption  and (stimulated) emission rates render it non-zero.
As a result, the only contribution in Eq.~({\ref{responsegeneral}}) comes 
from modes $n=\rightarrow$ [see Eq.~(\ref{n})]. Using the fact that 
$
|{{\cal A}^{(n){\rm abs}}_{\omega lm}}|
=
|{{\cal A}^{(n){\rm em}}_{\omega lm}}|
$,
we write Eq.~(\ref{totalresponsegeneral}) as 
\begin{equation}
     R_S \equiv 
     \frac{1}{\tau}
     \sum_{l=0}^{\infty} 
     \sum_{m} 
     \int_0^{\infty} d\omega   
       |{{\cal A}^{(\rightarrow){\rm em}}_{\omega lm}}  |^2 
       \coth (\omega \beta/2) \;.
    \label{totalresponsegeneral2}
\end{equation}

In order to deal with zero-energy modes, we need a ``regulator'' 
to avoid the appearance of intermediate indefinite results~\cite{HMS}.
For this purpose we let the  coupling constant $q$ to smoothly 
oscillate with frequency $\omega_0$, writing Eq.~(\ref{j(x)}) in 
the form [see Ref.~\cite{DS} for an alternative (but equivalent) 
regulator]
\begin{equation}
       j_{\omega_0}(x^\mu) = 
         ({ q_{\omega_0}}/{\sqrt{|h| \;}}) \delta(r-r_0) 
\delta^p (\theta_i-{\theta_{i}}_0) \;,
\label{j(x)2}
\end{equation}
where $ q_{\omega_0} \equiv \sqrt{2} q \cos(\omega_0 t)$ 
and  take the limit 
$\omega_0 \rightarrow 0$ at the end. The factor $\sqrt{2}$ 
has been introduced to guarantee that the time average
$\langle \; |q_{\omega_0}(t)|^2 \rangle_t$ equals $q^2$. 
By using Eqs.~(\ref{expansion}), (\ref{j(x)2}) and (\ref{SI}), 
we obtain  
\begin{equation}
|{\cal A}^{(\rightarrow) {\rm em}}_{\omega l m} |^2 = 
                    2 \pi^2 q^2 f(r_0) r_0^{-p} 
                    |\varphi^{(\rightarrow)}_{\omega_0 l} |^2 
                    | Y_{l m} |^2 [\delta(\omega - \omega_0)]^2 \;.
\label{emissionamplitude}
\end{equation}

Now we proceed to find the zero-energy modes with which our static
source interacts. For this purpose we let
$\omega = 0$  in Eq.~(\ref{Rwdif0}) and make the
change $ r \mapsto z \equiv 2(r/r_H)^{p-1}-1 $, obtaining 
\begin{equation}
(1-z^2) \frac{d^2 \psi^{(n)}_{0 l}}{dz^2} -2z \frac{d \psi^{(n)}_{0 l}}{dz}
+\frac{l(l+p-1) }{(p-1)^2} \psi^{(n)}_{0 l}  = 0 \; ,
\label{Eqomega0}
\end{equation}
where $1 < z < \infty$.
Two linearly independent solutions of 
Eq.~(\ref{Eqomega0}) can be given as
\begin{equation}
P_{\nu}(z) = F(-\nu, \nu + 1; 1; (1-z)/{2}) \; ,
\label{P}
\end{equation}
and
\begin{equation}
Q_{\nu}(z)\! = \frac{\Gamma(\nu +1) \Gamma \left({1}/{2} \right)}
                  {(2z)^{\nu +1} \Gamma \left( \nu + \frac{3}{2} \right)} 
F\left( \frac{\nu +2}{2}, \frac{\nu +1}{2}; \frac{2\nu +3}{2}; 
\frac{1}{z^2} \right)
\label{Q}
\end{equation}
with $\nu = l/(p-1)$. 
{}From the asymptotic behavior of the Legendre functions,
$ Q_{\nu}(z) \approx z^{-\nu -1} $ for $z \to \infty$
and
$ P_{\nu}(z) \approx 1$ for $z \approx 1$, 
we infer that, for $\omega \approx 0$,
\begin{eqnarray}
&& \varphi^{(\rightarrow)}_{\omega l} 
\approx   
C_{\omega l}^{(\rightarrow)} r^{p/2}    Q_{l/(p-1)}(z) \, ,   
\label{phi0}
\\
&& \varphi^{(\leftarrow)}_{\omega l} 
\approx
C_{\omega l}^{(\leftarrow)} r^{p/2}    P_{l/(p-1)}(z)  
\label{phi0'}
\end{eqnarray}
with 
$C_{\omega l}^{(n)} $ 
being normalization constants, generalizing a result with $p=2$ in
Ref.~\cite{candelas}. 
Now, by using 
Eqs.~(8.822.2) and~(3.513.2) of Ref.~\cite{GR}, and 
$
x  \approx  [{r_H}/{(p-1)}] \ln ( r/r_H -1 )\ \ {\rm for}\ \ r\approx r_H
$,
we obtain for $ x\to -\infty$ with $|\omega x| \ll 1$
\begin{equation}
\varphi^{(\rightarrow)}_{\omega l} (z) \approx 
-(  C^{(\rightarrow)}_{\omega l}r^{p/2}   /2) \ln (r/r_H-1) 
\,.
\label{phicloserh2}
\end{equation}
In order to find $C_{\omega l}^{(\rightarrow)} $, we firstly note from 
Eq.~(\ref{51a}) that close to the horizon and for small enough 
frequencies ($ x\to -\infty$, $|\omega x| \ll 1$):
\begin{equation}
\varphi^{(\rightarrow)}_{\omega l} 
\approx 
(4 \pi \omega)^{-1/2} 
[          
           (1+{\cal R}_{\omega l}^{(\rightarrow)}  ) + 
i \omega x (1- {\cal R}_{\omega l}^{(\rightarrow)} )   
] \;.
\label{phicloserh}
\end{equation}
Now, by comparing
Eqs.~(\ref{phicloserh2}) and~(\ref{phicloserh}), we note that
$
{\cal R}_{\omega l}^{(\rightarrow)} 
        \longrightarrow
  -1 
$
as $\omega \to 0$ and 
$$
C^{(\rightarrow)}_{\omega l} = 
{-2 i \sqrt{{\omega}/{\pi}\,} \, r_H^{1-p/2}}/{(p-1)} \,,
$$ 
which allows us to write for $\omega\approx 0$ [see Eq.~(\ref{phi0})] 
\begin{equation}
\varphi^{(\rightarrow)}_{\omega l} (z) \approx 
-2i \frac{\sqrt{\omega/\pi}}{p-1} \frac{r^{p/2}}{r_H^{p/2-1}} 
Q_{{l}/{(p-1)}}(z) \, .
\label{phi0final}
\end{equation}

Next, using Eq.~(\ref{phi0final}) in Eq.~(\ref{emissionamplitude}) 
and letting $\omega_0 \to 0$, we can write the response 
(\ref{totalresponsegeneral2}) as 
\begin{equation}
R_S = \frac{8 q^2 f^{1/2} (r_0) r_H^{2-p}}{\beta (p-1)^2} 
      \sum_{l=0}^{\infty} {Q_{l/(p-1)} (z_0)}^2
      \sum_{m}  | Y_{l m} |^2 \, ,
\label{totalresponsegeneral3}
\end{equation}
where $z_0 \equiv 2(r_0/r_H)^{p-1} -1 $. 
Using now
$$
\sum_{m}  | Y_{l m} |^2  = 
\frac{ (2l+p-1)(p+l-2)! \Gamma[(p+1)/2] }{{2 \pi^{(p+1)/2} (p-1)!\, l! } } \;,
$$
we eventually have
\begin{eqnarray}
&& R_S = \frac{q^2 f^{1/2} (r_0) r_H^{1-p} 
\Gamma[(p+1)/2]}{(p-1)  (p-1)!\, \pi^{(p+3)/2} } 
\sum_{l=0}^{\infty}
\left[ 
\frac{(2l+p-1) }{ l!}
\right.
\nonumber \\
& & 
\left. 
\times (p+l-2)! \left[ Q_{{l}/(p-1)} ( z_0 ) \right]^2 
\frac{}{} 
\right]\; .
\label{RS}
\end{eqnarray}
\begin{figure}
\begin{center}
\epsfig{file=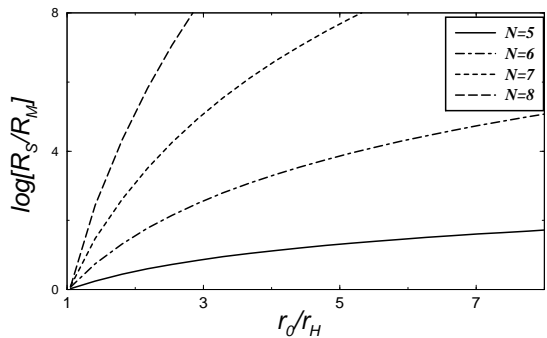,scale=.34,angle=-90}
\caption{\footnotesize Plot of the logarithm of the ratio $R_S/R_M$ 
as a function of
the source's radial coordinate $r_0$ (normalized by the Schwarzschild 
radius $r_H$) for $N=5, 6, 7, 8$. The source is assumed to have 
the same proper acceleration $a_0$ in Minkowski as in Schwarzschild 
spacetime. 
}
\label{Fig01}
\end{center}
\end{figure}
The expression above will be compared with the total response of
the source when it is uniformly accelerated in $N(=p+2)$ dimensional 
Minkowski spacetime with the line element 
$
ds^2 = dt^2 - dx^2 - d{\bf x}_\bot^2
$
({\it i.e.}, static in the corresponding Rindler wedge), 
where ${\bf x}_\bot = (x_2, \dots, x_{p+1})$. We assume that our 
source $j (x^\mu)$ is uniformly accelerated along  the $x$ axis 
with proper acceleration $a_0$ and is coupled to the scalar field 
$\hat \Phi (x^\mu) $ through the interaction action~(\ref{SI}). 
Here
$ 
    \hat{\Phi}(x^\mu) = 
        \int d k
        \int d {\bf k}^p_\bot 
        \left[ 
        u_{k {\bf k}_\bot} \hat{a}_{k {\bf k}_\bot} + {\rm H.c.} 
        \right]
$,
where
$
u_{k {\bf k}_\bot} = (2\omega (2\pi)^{n-1})^{-1/2} e^{- i k_\mu x^\mu} 
$, 
$k^\mu = (\omega, k, {\bf k}_\bot ) $, 
$\omega = \sqrt{k^2 + {\bf k}_\bot^2 }$
and  
$
[ \hat{a}_{k {\bf k}_\bot}, \hat{a}_{k' {{\bf k'}_\bot}}^\dagger ] = 
\delta(k-k' ) \delta^{p}( {{\bf k}_\bot} - {{\bf k'}_\bot}  )
$.
The total response in the Minkowski vacuum is 
$     
          R_M \equiv  
          \tau ^{-1}
          \int d k \int d{\bf k}^{p}_\bot  
          |{\cal A}^{em}_{k {\bf k}_\bot}|^2  
$,
where $\tau$ is the source's total proper time and
$
{\cal A}^{\rm em}_{k {\bf k}_\bot} 
\equiv 
\langle k \, {\bf k}_\bot | \hat S_I | 0 \rangle
$.
After integrating over the momentum $k$, 
we find for $N \geq 3$ 
\begin{equation}
R_M = \frac{2 q^2}{(2\pi)^{n-1} \, a_0}  \int d{\bf k}^{p} _{\bot} 
[K_0 ( {{\it k}_{\perp}}/{a_0} ) ]^2 \,
\, ,
\label{RMpre}
\end{equation}
where $k_\bot = |{\bf k}_\bot|$.
Now, using Eq.~(6.576.4) of Ref.~\cite{GR} and
$
\int d {\bf k}^p_{\bot} = 
\int_{0}^{\infty} d k_\bot k_\bot^{p-1}
\int d\Omega_{p-1} 
$
for $p \geq 2$,  
where $d\Omega_{p-1}$ is the volume element of the 
unit $p-1$ sphere, we perform the integration in Eq.~(\ref{RMpre})
(for $p=1$ the integration is trivial):
\begin{equation}
R_M = 
\frac{q^2 a_0^{p-1} \left[ \Gamma \left( {p}/{2} \right) \right]^4 
     \Omega_{p-1}}{8 \pi^{p+1}\, (p-1)!} 
\,,
\label{RM}
\end{equation}
where 
$
\Omega_{m} = {2\, \pi^{(m+1)/2}}/{\Gamma \left[ {(m+1)}/{2} \right]} 
$
for $m \geq 1$, and $\Omega_m =1$ for $m=0$. (See Ref.~\cite{takagi}
for related calculations.) 

For $N=4$ the responses~(\ref{RS}) and~(\ref{RM}) can be shown analytically 
to be identical [and to satisfy Eq.~(\ref{equivalence})], by using 
the equation $\sum_{l=0}^{\infty} (2l+1) [Q_l (z)]^2 = 1/(z^2 -1)$.
For $N \geq5 $, we were only able to compare numerically  the 
responses~(\ref{RS}) and~(\ref{RM}) (see Fig.~\ref{Fig01}). We first note 
that $R_S/R_M \approx 1$  for $r \approx r_H$ for every dimension 
$N \geq 4$. This is expected (see Ref.~\cite{HMS2}) and can be  
seen as a 
consistency check for our results. It is also clear from the graph 
that the full equality $R_S = R_M $ found in \cite{HMS2} is not valid 
for $N\geq 5$. This is the main result of the paper. 
It may be that
Eq.~(\ref{equivalence}) turns out to be 
a ``coincidence'' rather than a result of a deep principle yet to be
discovered. However, it is
worthwhile to note that this remarkable relation
appears precisely
in  spacetimes with the number of (macroscopic) dimensions of our 
physical world.

\begin{acknowledgments} 

G.M. is thankful to Funda\c c\~ao de Amparo \`a Pesquisa do Estado de 
S\~ao Paulo and  Conselho Nacional de Desenvolvimento Cient\'\i fico 
e Tecnol\'ogico for partial support.

\end{acknowledgments} 

\end{document}